\documentclass{iopjournal}

\usepackage{amsmath, amssymb}
\usepackage{array}
\usepackage{float}

\newcommand{\tr}{\mathrm{tr}}

\begin{document}
\articletype{Paper}

\title{Generation of arbitrary quantum states in cavity-atom systems with a three-level ancilla}

\author{Da-Wei Luo$^{1,\dagger}$\orcid{0000-0001-8226-5421}, Ting Yu$^{1,*}$}

\affil{$^1$Center for Quantum Science and Engineering and Department of Physics, Stevens Institute of Technology, Hoboken, New Jersey 07030, USA}

\email{${}^\dagger$dawei.luo@stevens.edu, ${}^*$ting.yu@stevens.edu}

\date{\today}

\keywords{quantum cavity systems, quantum control, quantum optics, quantum entanglement}

\begin{abstract}
We generalized the seminal work of Law \& Eberly~\cite{Law1996a}, and propose a protocol to generate arbitrary quantum states with a three-level atom as an ancilla in a more efficient manner. This protocol moves two excitations in parallel in each control step, thus reducing the number of evolution steps by half. Full analytical expressions for the control field strengths and timings are derived. The system dynamics is exposed to less chance of error accumulation in the timing and amplitudes of the control fields, and quantitative study in the context of the quantum speed limit is also considered.
\end{abstract}

\section{Introduction}

It is a great honor to contribute this paper to the Focus Collection Celebrating the Life and Science of Joseph H. Eberly. His pioneering research in quantum optics, optical physics, and quantum information left an enduring impact on these fields. Equally meaningful was his role as a generous and inspiring mentor; his intellectual insight, encouragement, and dedication shaped the careers of generations of students and colleagues. We dedicate this work to his memory with deepest gratitude and respect.

Quantum state engineering lies at the heart of quantum technologies. Entangled state~\cite{Horodecki2009a,Nielsen2000a} generation is necessary for various quantum information tasks such as quantum teleportation~\cite{Bennett1993a,Pirandola2015a}, quantum communication and computation~\cite{Nielsen2000a}, and can also offer enhanced precision and sensitivity for quantum metrology tasks~\cite{Degen2017a,Giovannetti2006a,Pezze2018a}.
For quantum optical cavity systems, state engineering is also indispensable for exploring nonclassical phenomena and probe light-matter interactions~\cite{Agarwal2012a,Gerry2023a}.
Preparing specific cavity states can enable the observations of Rabi Oscillation~\cite{Brune1996a},
and Schr\"odinger cat or Gaussian states provide valuable resources for investigating quantum interference~\cite{Pan2023a}, as well as quantum cryptography quantum error correction~\cite{Braunstein2005a,Weedbrook2012a}.
More generally, engineering cavity states constitute an integral step in quantum simulation~\cite{Georgescu2014a}, realization of quantum gates~\cite{Weedbrook2012a}, and encoding quantum states~\cite{Vlastakis2013a}. These applications motivate efficient and analytical protocols for preparing arbitrary states in quantum cavity systems.

To realize the task of state engineering, various approaches have been studied, such as optimal controls~\cite{Konnov1999b,Muller2022a,Chen2025a} which aims to minimize some cost function, stimulated Raman adiabatic passage~\cite{Vitanov2017a} which couples the population transfer states with radiations fields via an intermediate state, and through wave mixing and amplified spontaneous emission~\cite{Agarwal1986a}. State synthesis by reservoir engineering has also been reported~\cite{Kienzler2015a}, where the use of quantum neural network has also been proposed~\cite{Ghosh2019a}.
In the seminal work~\cite{Law1996a}, a fully analytical and deterministic protocol to generate arbitrary cavity state is propose for a Jaynes-Cummings like system. By alternating between a quantum field channel which enables the transition between \(|n-1, e \rangle\) and \(|n, g \rangle\), where \(|n \rangle\) is the cavity Fock state and \(e(g)\) are the excited (ground) state of a two-level atom, and a classical field channel that enables the transitions between \(|e\rangle\) and \(|g \rangle\), it was designed such that one photon can be removed from an arbitrary target state in one such step in a unitary fashion. Repeating this step, photons can then be successively removed until one reaches a vacuum state, then inverting this process since every step is unitary, one can realize the generation of the given target state at a prescribed time from an initial vacuum state. This simple yet powerful idea has been extended to generate entangled state between atom-field and between multiple bosonic modes~\cite{Kneer1998a, Zheng2000a}, with a possible realization through a quantum circuit~\cite{Santos2005a}, where it was suggested that parallel processing may be possible. Sequential generation of entangled photonic and atomic multiqubit states has also been suggested~\cite{Schon2007a}. Using three-level system as the ancilla, a cavity Fock state generation protocol~\cite{Santos2001a} has been proposed where the generation is conditional upon a measurement on the atom. Engineering two-mode interactions has also been shown to be possible by using three-level system interacting with two vibrational modes~\cite{Steinbach1997a}. Three-level atom ancilla has also been used for coherent state generation via repeated atom-field interactions~\cite{Plastina1999a}, and generating entanglement between two cavity modes~\cite{Ikram2002a}. A protocol to generate \(N\)-photon Fock states from \(N\) three-level systems has also been proposed~\cite{Brown2003a}. Here, we propose a generalization of the protocol~\cite{Law1996a} by using a three-level system as ancilla, in order to make the original protocol more efficient, by cutting down the number of control step needed to generate an arbitrary state with \(N\) excitations, and probe its efficiency via quantum speed limits (QSL)~\cite{Campo2013a,Taddei2013a,Toth2014a,Nishiyama2026a}, where there exists a minimal time needed to incur a change in the state's fidelity with respect to some reference state, or equivalently, the speed of fidelity change is both upper- and lower-bounded. This paper is organized as follows. We first outline our protocol in Sec.~\ref{sec_prot}, and then proceeds to discuss its efficiency and speed limits in Sec.~\ref{sec_rbnqsl}. Entangled states between the qutrit and cavity is also considered in Sec.~\ref{sec_entg}. The analytical derivations of the control field strengths and quantum speed limits are outlined in the appendix.

\section{Generation of cavity states with a three-level ancilla} \label{sec_prot}

To generate an arbitrary cavity state, one powerful analytical strategy proposed in the seminal paper~\cite{Law1996a} involves using a two-level system (TLS) as an ancilla, and by working backwards in time, one can piecewise apply a quantized field channel associated with the Hamiltonian \(\sigma_+ a + h.c.\) and a classical field channel associated with the Hamiltonian \(\sigma_+ + h.c.\) alternatively to remove one photon from the cavity state successively from the target state in a unitary fashion, until all photons are exhausted, which is the initial vacuum state for the state preparation protocol. It was later generalized such that entangled states between the atom and field, as well as multi-mode entangled states may be generated using similar ideas~\cite{Kneer1998a}, and a quantum gate based realization has also been discussed~\cite{Santos2005a}. Here, we ask the question: can this process be made more efficient, while keeping the process analytically solvable?

Consider the model of a \(\Delta\)-type three-level system interacting with an optical cavity, with a Hamiltonian of the form~\cite{Law1996a,Kneer1998a}, in the interaction picture,
\begin{align}
    H_{int} &= \sum_j H_{C}^{(j)} + H_{Q}^{(j)}, \nonumber \\
    H_{C}^{(j)} &= | j \rangle\langle j| \otimes \left[r_{1,j}  J_{m,g} + r_{2,j} J_{e,m} + r_{3,j} J_{e,g}\right]  + h.c., \nonumber \\
    H_{Q}^{(j)} &= g_{1,j} |j-1 \rangle\langle j| J_{m,g} + g_{2,j} |j-2 \rangle\langle j-1| J_{e,m} + h.c., \label{eq_h_3l}
\end{align}
where we denote \(J_{a,b} = | a \rangle\langle b |\) for \(a,b \in \{g,m,e\}\) for the three-level system, \(|n \rangle\) as the Fock basis for the cavity, and \(H_{C}\)(\(H_{Q}\)) describes the classical(quantized) field channel. The quantized field channel conserves the total number of excitations \(N = \sum_n n |n \rangle \langle n | + J_{mm} + 2 J_{ee}\), while the classical field channel keeps the cavity state unchanged and acts on the coherent transitions between the 3 ancilla levels \(g,m,e\) for ground, middle, and excited levels. Here, we switched the two-level system (TLS) in the model proposed in~\cite{Law1996a,Kneer1998a} with the \(\Delta\)-type three-level system, and keeps the field annihilation \(\propto |k-1 \rangle\langle k|\) distinct in each \(n\)-excitation subspace as~\cite{Kneer1998a}. This approach allows more degrees of control than using an annihilation operator for the cavity-atom coupling, and can, crucially, allow us to design a protocol that reduce the number of steps for the state preparation by a factor of \(2\).

\begin{figure}
    \centering
    \includegraphics[width=.85\textwidth]{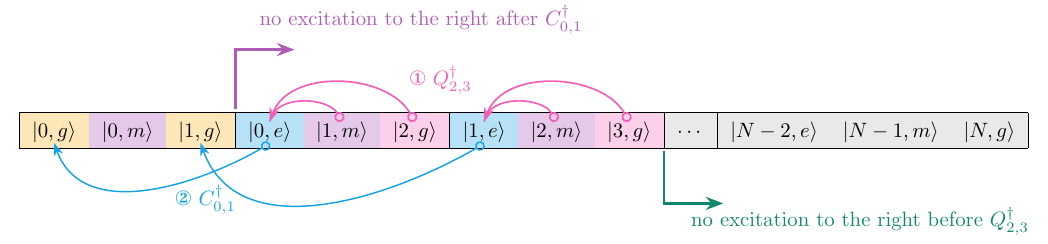}
    \caption{Schematic of the state preparation protocol using a three-level system coupled to a cavity. The unitary \(Q^\dagger\) steps conserves the total number of excitations and moves the wave function elements to the \(|n-2, e \rangle, |n-3, e \rangle\) positions for blocks with \(n\) and \(n-1\) excitations in each step. The unitary \(C^\dagger\) steps coherently transfers them to their respective atomic \(|g \rangle\) states while keeping cavity photon number unchanged. In each reversal cycle, we first apply the unitary \(Q_{j,j+1}^\dagger\) step to move the states in the subspace with \(j\) and \(j+1\) excitations such that the atom is in \(|e \rangle\) in the subspace. We then apply the unitary \(C_{j-2, j-1} ^\dagger\) step to transfer the atomic state from \(|e \rangle\) to \(|g \rangle\) without touching the photon numbers. Repeating this process, we can work backwards to unitarily map any quantum state back to the \(|0,g\rangle\) initial state, such that the state preparation can be done with piecewise alternating \(C\), \(Q\) steps.}\label{fig_schema}
\end{figure}

The schema for the proposed protocol is illustrated in Fig.~\ref{fig_schema}. Note that while \(H_C\) and \(H_Q\) does not commute, each \(H_{Q}^{(j)}\) only acts on subspaces with \(j\) excitations, and similarly, each \(H_{C}^{(j)}\) only acts on subspaces with \(j\) photons in the cavity state so we have the commutation relationship
\begin{align}
    [H_{Q}^{(i)}, H_{Q}^{(j)}] = 0, \;
    [H_{C}^{(i)}, H_{C}^{(j)}] = 0. \label{eq_comu}
\end{align}
The goal of the protocol is to generate a target state at the end of a given runtime \(T\),
\begin{align}
    |\psi(T) \rangle = \sum_{n=0}^{N}\sum_{i=g,m,e} c_{n, i} |n, i \rangle,
\end{align}
from the initial state \(|\psi(0) \rangle = |0, g \rangle\), where \(C_{N,m(e)}=C_{N-1,e}=0\) to limit target states to a subspace with at most \(N\) total excitations.

\begin{figure}
    \centering
    \includegraphics[width=.7\textwidth]{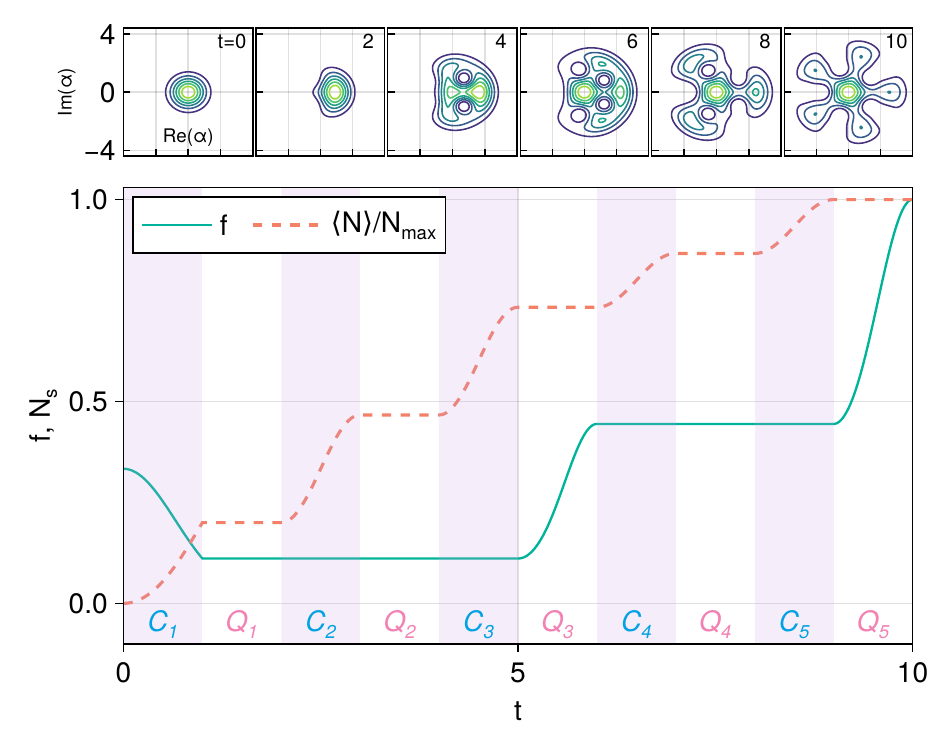}
    \caption{Fidelity (green solid line) and scaled excitation (coral dashed line) \(N_s = \langle N \rangle / N_{\rm max}\) dynamics, with snapshots of the Husimi Q functions at the end of each \(C\)-\(Q\) cycle. Due to the design of this protocol and the target state considered here, the fidelity first backtracks then moves towards the target \(f=1\) at the end of the runtime, while the number of excitations are kept constant in each \(Q\) step and rises steadily in each \(C\) step.}\label{fig_step}
\end{figure}

Following similar ideas in~\cite{Law1996a,Kneer1998a}, we divide \([0,T]\) into $2m$ equal segments of length \(\tau=T/2m\), where in each segment, only \(H_C\) or \(H_Q\) is active, alternatively. The commutation relationship Eq.~\eqref{eq_comu} allows us to act on two consecutive blocks in Fig.~\ref{fig_schema} in parallel in each \(C^\dagger\) or \(Q^\dagger\) step since they have different even-odd parity of the number of total excitations. Let
\begin{align}
    Q_{i,j} = \exp\left[-i( H_{Q}^{(i)} + H_{Q}^{(j)} )\tau\right],\; C_{i,j} = \exp\left[-i( H_{C}^{(i)} + H_{C}^{(j)} )\tau\right],
\end{align}
and denote the single subscript symbols \(Q_m=Q_{N, N-1},\, Q_{m-1}=Q_{N-2, N-3},\, \ldots\) where \(N\) is the largest number of excitations in the target state, and likewise for \(C_m=C_{N-2, N-3}, \ldots\). The full unitary evolution propagator can then be given as \(U = Q_m C_m Q_{m-1}C_{m-1}\ldots Q_1 C_1\). Each \(C,Q\) block can be given analytically (see Appendix~\ref{sec_app_analy}). We then iteratively destroy 2 photons in 2 consecutive blocks simultaneously, until all excitations are exhausted, to arrive at \(|0, g \rangle = U^\dagger |\psi(T) \rangle\). Working backwards in time, we may first apply \(Q_j^\dagger\equiv Q_{n, n-1}^\dagger\) for the two subspaces with \(n\) and \(n-1\) excitations, to send
\begin{align}
    |\psi_n \rangle &\equiv c_{n-2, e}| n-2, e \rangle + c_{n-1, m}| n-1, m \rangle + c_{n, g}| n, g \rangle \quad\xrightarrow{Q_j^\dagger}\quad c_{n-2, e}' | n-2, e \rangle, \nonumber \\
    |\psi_{n-1} \rangle &\equiv c_{n-3, e}| n-3, e \rangle + c_{n-2, m}| n-2, m \rangle + c_{n-1, m}| n-1, g \rangle \quad\xrightarrow{Q_j^\dagger}\quad c_{n-3, e}' | n-3, e \rangle,
\end{align}
which involves 4 equations
\begin{align*}
    \langle n-1, m| V_j \rangle &= 0, \\
    \langle n, g| V_j \rangle &= 0, \\
    \langle n-2, m| V_j \rangle &= 0, \\
    \langle n-1, g| V_j \rangle &= 0, \\
\end{align*}
where
\begin{align}
    |V_j\rangle = Q_j^\dagger \left[|\psi_n \rangle + |\psi_{n-1} \rangle\right],
\end{align}
with 4 variables \(g_{1(2),n(n-1)}\) and always have solutions, where for the last reversal step it may need to send
\begin{align}
    |\psi_0 \rangle &\equiv c_{0, m}| 0, m \rangle + c_{1, g}| 1, g \rangle \quad\xrightarrow{Q_1^\dagger}\quad c_{0, m}' | 0,m \rangle,
\end{align}
with the equation
\begin{align}
    \langle 1, g | Q_1^\dagger |\psi_0 \rangle = 0,
\end{align}
for the variable \(g_{1, 1}\), see Appendix~\ref{sec_app_analy} for details. Then, in we apply \(C_j^\dagger\equiv C_{n-2, n-3}^\dagger\) to send
\begin{align}
    |v_{n-2} \rangle &\equiv c_{n-2, e}' | n-2, e \rangle + c_{n-2, g} | n-2, g \rangle \quad\xrightarrow{C_j^\dagger}\quad \tilde c_{n-2, g} | n-2, g \rangle, \nonumber \\
    |v_{n-3} \rangle &\equiv c_{n-3, e}' | n-3, e \rangle + c_{n-3, g} | n-3, g \rangle \quad\xrightarrow{C_j^\dagger}\quad \tilde c_{n-3, g} | n-3, g \rangle,
\end{align}
which involves 2 equations
\begin{align*}
    \langle n-2, e | w_j \rangle &= 0, \\
    \langle n-3, e | w_j \rangle &= 0,
\end{align*}
where
\begin{align}
    |w_j\rangle = C_j^\dagger \left[|v_{n-2} \rangle + |v_{n-3} \rangle\right],
\end{align}
with 2 variables \(r_{3, n-2(n-3)}\). Note for the last reversal step, depending on the target state or parity of \(N\), we may need to send
\begin{align}
    |v_{0} \rangle &\equiv c_{0, e} | 0, e \rangle + c_{0, m} | 0, m \rangle + c_{0, g} | 0, g \rangle \quad\xrightarrow{C_1^\dagger}\quad c_{0, g}' | 0, g \rangle,
\end{align}
which involves 2 equations
\begin{align*}
    \langle 0, e | w_1 \rangle &= 0, \\
    \langle 0, m | w_1 \rangle &= 0,
\end{align*}
for
\begin{align}
    |w_1 \rangle = C_1^\dagger |v_{0} \rangle,
\end{align}
for 2 variables \(r_{1(2), 0}\).

\begin{figure}
    \centering
    \includegraphics[width=.7\textwidth]{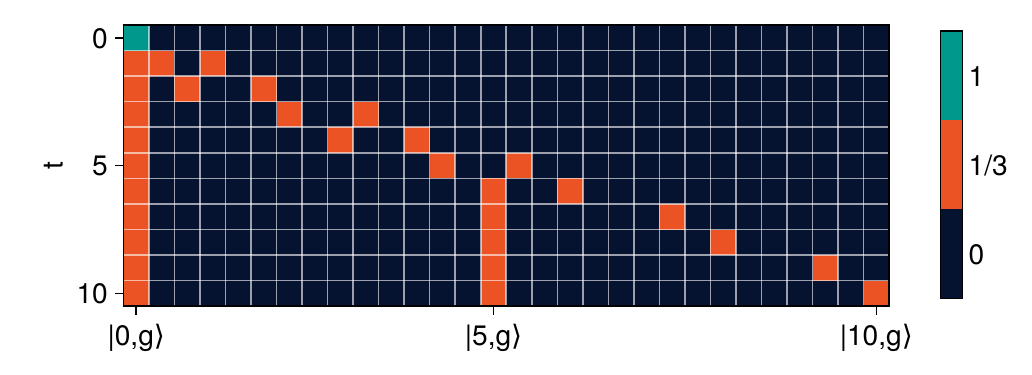}
    \caption{Wave function vector element \(|c_{n,i}|^2\) at the end of each control step: y axis is the evolution time, and the basis is arranged as dressed states in the order specified in Fig.~\ref{fig_schema} for the x axis.}\label{fig_psi_elem}
\end{figure}

As an illustrative example, we consider the target `flower' cavity state~\cite{Law1996a},
\begin{align}
    |\psi_{\rm tgt} \rangle = \left( |0 \rangle + |5 \rangle + |10 \rangle \right) \otimes |g \rangle / \sqrt{3}.
\end{align}
This target has a maximum of \(10\) photons, which would need 10 \(C\text{-}Q\) steps to realize with a TLS ancilla. With a three-level ancilla Eq.~\eqref{eq_h_3l}, we show that this state can be realized in half the number steps, with analytical solutions for the field strengths (see Appendix~\ref{sec_app_roots} for details).

Setting \(\tau=1\), the dynamics of the fidelity \(f = |\langle \psi(t) | \psi_{\rm tgt} \rangle|^2\) is shown in Fig.~\ref{fig_step}, with snapshots the of Husimi Q-function
\begin{align}
    Q(\alpha) = |\langle \alpha | \rho_c(t) | \alpha \rangle|,
\end{align}
where \(|\alpha \rangle\) is the coherent state and we denote \(\rho_c = \tr_{a} |\psi(t) \rangle\langle \psi(t) |\) is the reduced density operator for the cavity after tracing out the ancilla degree of freedom. It can be noted that the \(Q\) steps conserves the total number of excitations, while the \(C\) step steadily increases it to the target as the state evolves. Also due to the definition of the fidelity as an inner product, we do not get any credit for moving a wave function vector elements closer to its index location - the inner product of \(|10 \rangle\) with \(|1 \rangle\) and \(|9 \rangle\) are both zero, even though with this protocol moving to the Fock state \(|9 \rangle\) is conceptually closer to the target state. Therefore, in the dynamics of fidelity, there are periods of constant fidelity but the protocol is still functional and drives the system to its target state at the prescribed runtime \(T\). We also visualize how the vector element of the wave function are being moved to their target positions at each control step in Fig.~\ref{fig_psi_elem}. At the end of the first \(C_1\) step \(t=1\), the Hamiltonian drives the initial \(|0, g \rangle\) state to an equal weight state of \(\left[ |0, g \rangle + |0, m \rangle + |0, e \rangle \right]/\sqrt{3}\). The \(|0, g \rangle\) component is already in the target position, and subsequent steps moves the other 2 elements to their correct positions successively during the evolution.

\section{Robustness and efficiency of the protocol} \label{sec_rbnqsl}

In practical realizations, errors in both the timing and amplitude of the control fields are possible. Since the protocol discussed here using a three-level system can cut down the number of steps by half, one would expect the error accumulations would be lower. To study the robustness of the protocol, we introduced \(8\%\) error in both the field strength and timing of the \(Q\)-\(C\) steps for the protocols using TLS and three-level system as the ancilla.

\begin{figure}
    \centering
    \includegraphics[width=.65\textwidth]{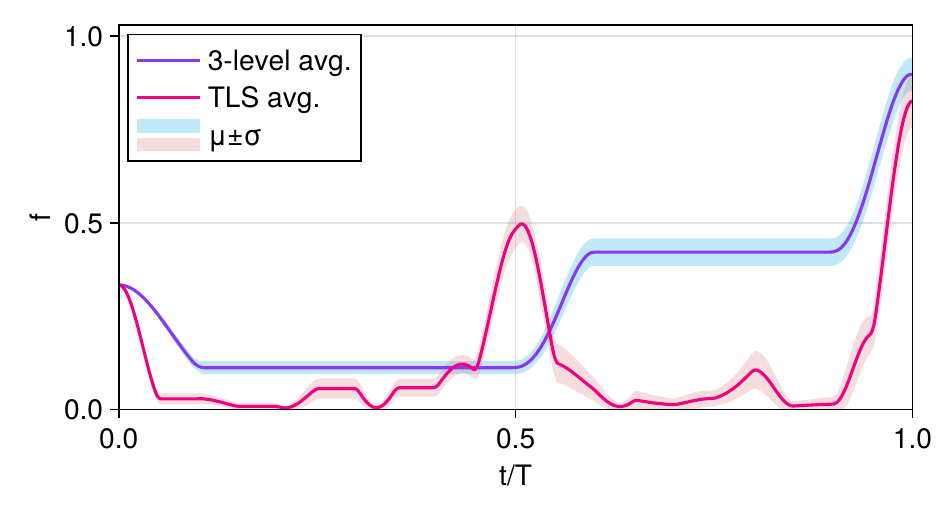}
    \caption{Fidelity dynamics after introducing \(8\%\) error in both the timing and amplitude of the control fields, \(10,000\) random samples, at a rescaled time \(t/T\) where \(T\) is the total runtime for using a TLS and a three-level atom as the ancilla. Since 3-level ancilla can reduce the number of steps needed to generate the same target state, it accumulates less error and achieve a higher final fidelity \(f \approx 0.90\) compared to \(f \approx 0.83\) for the TLS. Shaded area indicates the error spread of the fidelity dynamics for the random samples as indicated by \(\mu\pm \sigma\), where \(\mu\), \(\sigma\) is the mean and standard deviation of the samples at each time point respectively. It can be seen that the three level case has less error spread than the TLS case.}\label{fig_err}
\end{figure}

Taking \(10,000\) random samples for the error analysis, we plotted the results in Fig.~\ref{fig_err}. Since the three-level atom ancilla protocol needs only half the number of steps, it has less chance to accumulate errors, and the resultant final fidelity is much higher than the TLS case. We also calculated the relative standard deviation \(\eta = \sigma/\mu\) for the two cases, where \(\sigma\) is the standard deviation and \(\mu\) is the mean. For the three-level case, we have \(\eta \approx 5.0\times 10^{-2}\) while the TLS case gives a bigger \(\eta \approx 8.6\times 10^{-2}\), which is also observable in Fig.~\ref{fig_err} where the TLS case's error spread is larger. This suggests that by using an ancilla of higher dimension, the state preparation protocol can be more robust to control errors in the field timing and amplitude.

On the other hand, quantum system dynamics is constrained by a quantum speed limit (QSL)~\cite{Campo2013a,Taddei2013a,Toth2014a,Nishiyama2026a}, where there exists a minimal time needed to incur a change in the state's fidelity with respect to some reference state, or equivalently, the speed of fidelity change is upper- and lower-bounded. For time-independent systems, such bound is given by the variance of the energy~\cite{Campo2013a,Nishiyama2026a},
\begin{align}
    T \geq \frac{\arccos \sqrt{f}}{\Delta H},
\end{align}
or, equivalently \(\arccos \sqrt{f} \leq \Delta H \times T\). In the time-dependent case, the RHS becomes an integral (see Appendix~\ref{sec_app_qsl} for details),
\begin{align}
    |\varphi(T) - \varphi(0)| &\leq \int_0^T ds \sqrt{\mathcal{F}(s)}/2 \equiv \Gamma, \label{eq_qsl_int}
\end{align}
or in the differential form,
\begin{align}
    |\dot \varphi|  &\leq \sqrt{\mathcal{F}}/2 \equiv \dot \Gamma, \label{eq_qsl_deriv}
\end{align}
where \(\varphi = \arccos \sqrt{f}\) and \(\mathcal{F}\) is the quantum Fisher information (QFI)~\cite{Liu2019a,Toth2014a} with time as its formal parameter. For system under unitary evolution, \(\Gamma\) reduces to the integral of the variance of energy, \(\Gamma =  \int_0^T ds \Delta H(s)\). The differential forms shows locally the rate of change of fidelity is bounded both up and below, while with the integral form, we can take \(\Gamma\) encoding both the evolution time and the variance of energy as a resource to affect a change in the quantum state's fidelity with respect to some reference state, which is set to the target state here. Since the main purpose of this protocol is to generate a given target cavity state, there are two ways to study the QSL: for the composite cavity\(+\)ancilla system, or the trace out the ancilla and study the reduced quantum state of the cavity only, which evolves non-unitarily.

\begin{figure}
    \centering
    \includegraphics[width=.8\textwidth]{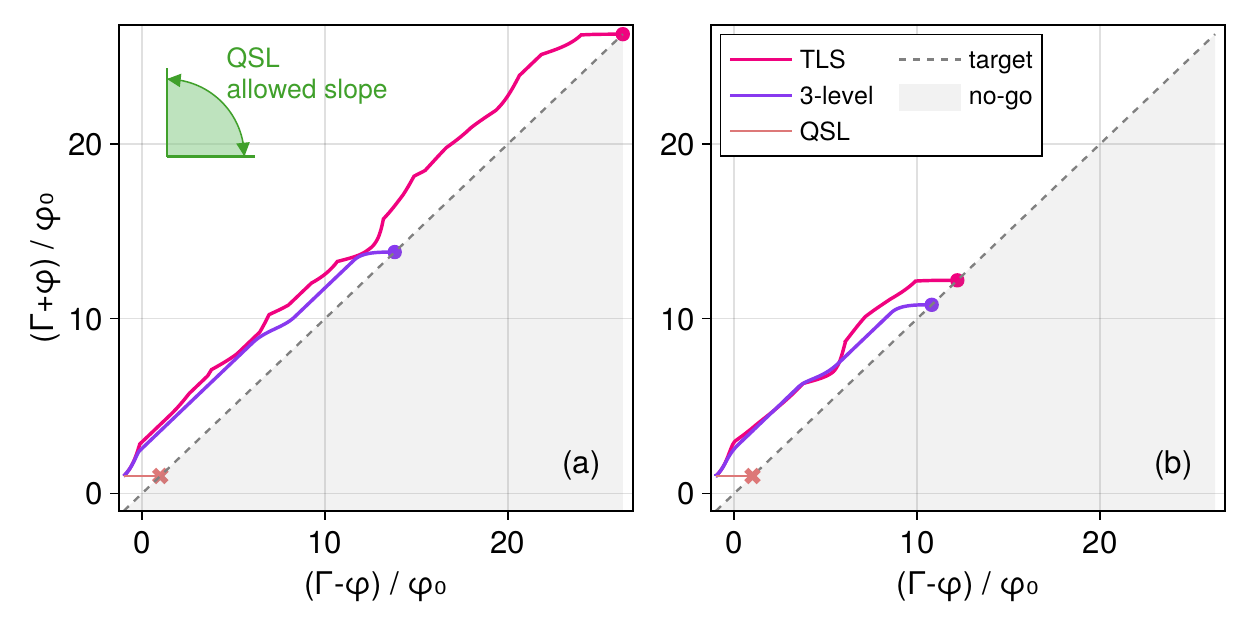}
    \caption{QSL comparison for protocols using three- ane two-level atoms as ancilla. Since \(\varphi \geq 0\), the grey shared area is the no-go zone, and horizontal movements of the trajectory signifies the saturation of QSL. State starts at \((-1, 1)\) and target is reached when it reaches the dashed \(y=x\) line. QSL limits the slope of the lines to be non-negative, where horizontal movement means the system is moving towards the target at QSL. Panel (a): the composite atom\(+\)cavity system, (b): the reduced cavity state as a non-unitary evolution process. Purple (magenta, orange) lines represents the three-level (TLS, ideal QSL geodesic) dynamics, with circle (cross) markers to signify the target is reached. It can be seen that for both protocols, they start out expending the the QSL resource \(\Gamma\) without gaining much fidelity, whereas it reaches the target state at the final step at near QSL efficiency.}\label{fig_qsl}
\end{figure}

We consider both cases and plots the QSL comparison in Fig.~\ref{fig_qsl} with \((\Gamma\pm \varphi)/\varphi_0\) as the \(y(x)\) axis, where \(\varphi_0\) is the fidelity angle between the initial and target state, \(\varphi_0 = \arccos |\langle \psi(0) | \psi_{\rm tgt} \rangle|\). Thus, states starts at coordinate \((-1, 1)\) and the target is reached when \(\varphi = 0\) which is the \(y=x\) line. Since \(\varphi \in [0, \pi/2]\), the area under the \(y=x\) line is not reachable.
The slope of the evolution is given by, when \(\dot \Gamma \neq \dot\varphi \),
\begin{align}
    \frac{dy}{dx} &= \frac{\dot\Gamma + \dot\varphi}{\dot\Gamma - \dot\varphi} = \frac{\dot\Gamma^2 - \dot\varphi^2}{(\dot\Gamma - \dot\varphi)^2},
\end{align}
which is restricted to be non-negative by the differential form of QSL, Eq.~\eqref{eq_qsl_deriv}. When the QSL is saturated, \(|\dot \varphi| = \dot \Gamma\), the slope is either \(0\) or infinity, corresponding to states moving towards or away from the target state at the speed limit.
In the plot, we can see that during the evolution, the lines are approximately parallel to the \(x=y\) line except at the end, where it moves almost horizontally. It would suggest that before the last evolution step, \(\Gamma\) as a resource is spent without affecting fidelity significantly, but at the final step it moves the state towards the target state almost at the limit imposed by QSL, which we can also verify in the dynamics plot Fig.~\ref{fig_step}. This effect is due to the definition of the fidelity as a simple inner product: consider the Fock state basis, this protocol moves the vector element to their target position incrementally. In this context, for example, the Fock state \(|7 \rangle\) may be conceptually closer to a target of \(|10 \rangle\) than the Fock state \(|1 \rangle\), but as far as fidelity is concerned, they are equally far away from the target state. In panel (b) of Fig.~\ref{fig_qsl}, we plot the QSL by tracing out the ancilla degree of freedom, where \(\Gamma\) is generalized to be an integral of QFI (see Appendix~\ref{sec_app_qsl} for details). The values of \(\Gamma\) are smaller after the partial trace, which can be understood as the sub cavity system contains less information than the composite cavity-ancilla system, and in both cases, we can see that by using three-level atom as the ancilla gives a trajectory that can be seen to be more efficient in the context of QSL and reaches the target state at lower costs as measured by \(\Gamma\) as a resource.

\section{Generating entangled cavity-atom states} \label{sec_entg}

\begin{figure}
    \centering
    \includegraphics[width=.7\textwidth]{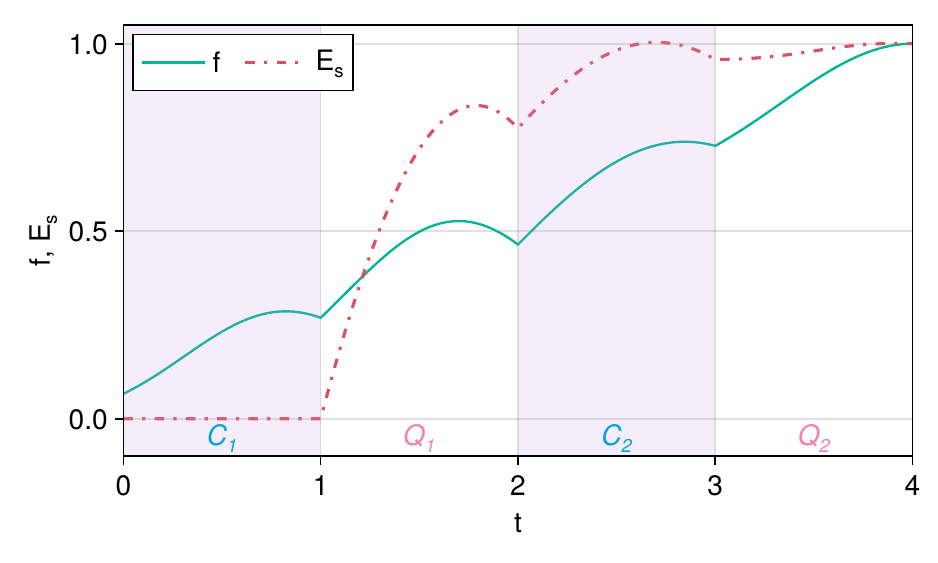}
    \caption{Generation of a sample entangled atom-cavity target state in 2 steps. Green solid line is the fidelity dynamics, and the red dot-dashed line represents the entanglement dynamics with a scaled log-negativity \(E_s = E/E_T\) where \(E_T\) is the target state entanglement.}\label{fig_ents}
\end{figure}

Since the protocol proposed relies on successively removing excitations, the target state is not limited to the form of \(|\psi_{\rm cav} \otimes |g \rangle\), and entangled states between the cavity and atom is also possible. As an example, we consider the entangled target state
\begin{align}
    |\psi_{\rm tgt} \rangle &=  \left[ |\psi_g \rangle |g \rangle + |\psi_m \rangle |m \rangle +|\psi_e \rangle |e \rangle \right] / {\sqrt{3}}, \nonumber \\
    |\psi_g \rangle &= \sum_{i=0}^4 |i \rangle / \sqrt{5},  \nonumber \\
    |\psi_m \rangle &= \left[|0 \rangle - |1\rangle + |3 \rangle \right] / \sqrt{3},  \nonumber \\
    |\psi_e \rangle &= \left[|0 \rangle + |1\rangle - |2 \rangle \right] / \sqrt{3},
    \label{eq_psiT_ent}
\end{align}
which has \(3\) non-zero Schmidt coefficients
\begin{align}
    c = \frac{1}{\sqrt{3}}, \, \frac{1}{3}\sqrt{\frac{15 \pm \sqrt{30}}{5}}.
\end{align}
Here, we only consider a pure state target prepared unitarily, so the entanglement may be measured by the logarithmic negativity~\cite{Vidal2002a,Plenio2005a} as
\begin{align}
    E = 2\log_2(\sum_\alpha c_\alpha),
\end{align}
where \(c_\alpha\) are the Schmidt coefficients.

Since the target state has a maximum of \(4\) photon, we are able to generate this entangled in \(2\) steps, where the coefficients for the Hamiltonian's field strength can be derived fully analytically (see Appendix~\ref{sec_app_roots} for details). We plot the entanglement and fidelity dynamics in Fig.~\ref{fig_ents}. We can see that this protocol is able to generate that target entangled state at the end of the prescribed runtime.
We would like to point out that while the \(C\) steps relates to the classical channel, \(H_C\) is not local to the cavity or ancilla and thus can induce changes in entanglement.
It is also worth nothing that this protocol is only design to reach the target state at the prescribed time, and no other restriction is imposed in the derivation of the controls. Therefore, the entanglement may reach or exceed that of the target state during evolution, which we may observe in the \(C_2\) stage of Fig.\ref{fig_ents}, where fidelity is lower than the target but the entanglement can actually be higher at certain times. Here, fidelity as a simple distance measure is not indicative of entanglement.

\section{Conclusion}
In conclusion, we build up on the seminal work~\cite{Law1996a, Kneer1998a} to generate arbitrary cavity or cavity-atom state in a more efficient manner, by considering a high-dimension ancilla. This setup allows us to remove multiple photons from the target state in parallel, and can therefore prepare the target state in less steps. Remarkably, the control field strengths can be fully derived analytically. The system would have less chances to accumulate errors in either the timing or the amplitude errors of the control fields, making it more robust to these imperfections. The efficiency is also quantitatively studied in the framework of quantum speed limits, and we find that using a three-level atom as ancilla gives a faster dynamics in terms of QSL. We also show that the protocol is versatile and speeding up the generation of entangled target states between the cavity and the ancilla is also possible. This protocol shows the power of simplicity of the ideas of Law and Eberly~\cite{Law1996a}, making it easily extensible to various generalizations and implementations.


\section{Acknowledgement}
This project is supported by ACC-New Jersey under Contract No. W15QKN-24-C-0004.

\appendix
\section{Appendix: analytical expressions for the evolution propagator and state preparation equations} \label{sec_app_analy}

Consider the classical channel, under the basis \( \{|n, g \rangle,\, |n, m \rangle,\, |n, e \rangle \} \), the effective Hamiltonian for the subspace is given by, when \(r_3 = 0\),
\begin{align}
    h = \begin{bmatrix}
        0     & r_1^* & 0 \\
        r_1   &   0   & r_2^* \\
        0     & r_2   & 0
    \end{bmatrix},
\end{align}
and \(u_C^\dagger = \exp [i h \tau]\) can be written as
\begin{align}
    u_C^\dagger = I + \frac{i \sin(\Omega\tau)}{\Omega} h + \frac{\cos(\Omega\tau) - 1}{\Omega^2} h^2,
\end{align}
where \(\Omega = \sqrt{|r_1|^2 + |r_2|^2}\). Also note that in the \(\{ | N, g \rangle,\, | N-1, m \rangle ,\, | N-2, e \rangle \}\) basis, the quantized field channel Hamiltonian also has this general structure, so the analysis below applies for \(Q\) and \(g_{1,2}\) too.

After some algebra, it can be shown that for the reversal step to satisfy
\begin{align}
    u_C^\dagger |\psi_{C0} \rangle \equiv u_C^\dagger \begin{bmatrix}
        \varphi_g \\
        \varphi_m \\
        \varphi_e
    \end{bmatrix} =
    \begin{bmatrix}
        \varphi_g' \\
        0 \\
        0
    \end{bmatrix}, \label{eq_c_fullsol}
\end{align}
the solution for \(r_{1,2} \tau\) can be given as, when \(\varphi_m \neq 0\)
\begin{align}
    r_1\tau &= i \frac{\varphi_m}{\gamma} \frac{\Omega}{\sin \Omega}, \label{eq_r1tau}\\
    r_2\tau &= i \frac{\Omega}{\eta} \frac{\varphi_e}{\varphi_m},
\end{align}
where
\begin{align*}
    \Omega &= 2 \arctan \eta,\,
    \gamma = \varphi_g/C_1, |\gamma| = \mathrm{norm}(|\psi_{C0} \rangle),\\
    C_1 &= \frac{1-\eta^2 + 2 |\varphi_e|^2/|\varphi_m|^2}{1+\eta^2},\\
    \eta^2 &= \frac{|\gamma| (1+2 |\varphi_e|^2/|\varphi_m|^2) - \sigma |\varphi_g|}{|\gamma| + \sigma |\varphi_g|}, \; \sigma=\pm 1.
\end{align*}
When \(|\varphi_g|=0\), \(|r_1 \tau|\) is given by Eq.~\eqref{eq_r1tau} and it can have a free phase.
When \(|\varphi_g|=|\varphi_e|=0\), we have \(r_2=0,\, r_1 \in R\) and \(\cos r_1\tau = 0\).

When \(\varphi_m = 0\), we may choose, if \(\varphi_g \neq 0\),
\begin{align}
    r_1\tau = \pi  \sqrt{\frac{|\gamma|  -| \varphi_g | }{2 |\gamma| }}, \quad
    r_2\tau = -\frac{\pi ^2 \varphi_e \varphi_g^*}{2 |\gamma \varphi_g | r_1},
\end{align}
If \(\varphi_g = 0\), we may either choose \(r_{1(2)}=0\) and \(r_3 = r_x e^{ir_y}\), any \(\cos(r_x)=0\) and \(ie^{-ir_y}\sin(r_x)=1\), one choice is
\begin{align}
    r_3\tau = i\pi/2,
\end{align}
or let \(r_3 = 0\) and
\begin{align}
    r_1 \tau = \pi / \sqrt{2}, \quad
    r_2 \tau = -\pi / \sqrt{2}.
\end{align}

Thus, for the protocol where we want to generate a cavity target state
\[
    |\psi(T) \rangle = \sum_{i=1}^{N} c_i |i \rangle \otimes |g \rangle,
\]
the full solutions to the controls are (depending on whether $N$ is even or odd, for the last reversal step):
\begin{enumerate}
    \item Set \(g_{1,j}\tau = g_{2,j}\tau = i\frac{\pi + 2\pi n}{\sqrt{2}}\) for any integer $n$, which flips $c_{j}|j,g \rangle$ to $c_{j}|j-2,e \rangle$ for 2 blocks: \(j=M\), \(j=M-1\) when \(j\geq 2\) for all but the last reversal step. For the last reversal step \(j=1\),  we have
    \(g_{2,j} = 0\), and
    \begin{align}
        i e^{i y_g} \sin (x_g) = 1,
    \end{align}
    where \(g_{1,j} = x_g e^{i y_g}\) which flips \(|1, g \rangle\) to \(|0, m \rangle\), one choice can be \(g_{1,j} = -i\pi/2\).
    \item Set \(r_{1,j} = r_{2,j} = 0\) for 2 blocks: \(j=M-2\), \(j=M-3\) when \(j > 0\) for all but the last reversal step,
    \(r_{3,j} = x_r e^{i y_r}\),
    \begin{align}
        x_r &= \arctan |\varphi_e/\varphi_g| \nonumber \\
        y_r &= \arg(\varphi_e/\varphi_g) + \pi/2 \label{eq_cg_sol}
    \end{align}
    when \(\varphi_g \neq 0\), otherwise \(x_r=y_r=\pi/2\). This flips $\varphi_e | j, e \rangle + \varphi_g | j, g \rangle \rightarrow \varphi_g' |j, g \rangle$, where
    \begin{align}
        \varphi_g' =  e^{i*\arg{\varphi_g}} \sqrt{|\varphi_g|^2 + | \varphi_e|^2}.
    \end{align}
    For the last step, depending on the target state or the parity of \(N\), either use Eq.~\eqref{eq_cg_sol} for \(r_{1,0} = x_r e^{i y_r}\), \(r_{2(3),0} = 0\) to flip $\varphi_m | 0, m \rangle + \varphi_g | 0, g \rangle \rightarrow \varphi_g' |0, g \rangle$, or use the solution for Eq.~\ref{eq_c_fullsol} to flip $\varphi_e | 0, e \rangle + \varphi_m | 0, m \rangle + \varphi_g | 0, g \rangle \rightarrow \varphi_g' |0, g \rangle$.
\end{enumerate}

\section{Appendix: solutions to the control fields} \label{sec_app_roots}

For the flower state \(\left( |0 \rangle + |5 \rangle + |10 \rangle \right) \otimes |g \rangle / \sqrt{3}\), the field strength are as follows, where parameter not explicitly given are all \(0\):
\begin{table}[H]
\centering
\renewcommand{\arraystretch}{1.4}
\newcolumntype{V}{>{\centering\arraybackslash}p{4cm}}
\begin{tabular}{|l|V||l|V|}
\multicolumn{4}{c}{flower state} \\
\hline
$Q_5$ & $g_{1(2), 10} = i\pi/\sqrt{2}$ & $C_5$ & $r_{3, 8} = i \pi / 2$ \\
\hline
$Q_4$ & $g_{1(2), 8}= i\pi/\sqrt{2}$ & $C_4$ & $r_{3, 6} = i \pi / 2$ \\
\hline
$Q_3$ & $g_{1(2), 6} = g_{1(2), 5} = i\pi/\sqrt{2}$ & $C_3$ & $r_{3, 4} = r_{3, 3} = i \pi / 2$ \\
\hline
$Q_2$ & $g_{1(2), 4} = g_{1(2), 3} = i\pi/\sqrt{2}$ & $C_2$ & $r_{3, 2} = r_{3, 1} = i \pi / 2$ \\
\hline
$Q_1$ & \rule[-3.5ex]{0pt}{8ex} $\begin{aligned} g_{1(2), 2} &= i\pi/\sqrt{2} \\ g_{1(2), 1} &= -i\pi/{2} \end{aligned}$ & $C_1$ & $\begin{aligned} r_{1, 0}\\ r_{2, 0} \end{aligned}$ \\
\hline
\end{tabular}
\end{table}
where \(s = \sqrt{5-2 \sqrt{3}}\) and
\begin{align}
    r_{1,0} &= \frac{2 i \arctan\left(s\right) \csc \left(2 \arctan s\right)}{\sqrt{3}} \approx 1.05363 i \\
    r_{2,0} &= \frac{2 i \arctan s}{s} \approx 1.43929 i
\end{align}

For the cavity-atom entangled target state Eq.~\eqref{eq_psiT_ent}, the parameters are analytically given as:
\begin{itemize}
    \item \(Q_2\):
    \begin{align*}
        g_{1, 4} &= -2 i \sqrt{\frac{3}{21-2 \sqrt{65}}} \arctan \left(\sqrt{\frac{1}{5} \left(21-2 \sqrt{65}\right)}\right) \\
        &\approx -1.22228i \\
        g_{2, 4} &= 2 i \sqrt{\frac{5}{13}} \arctan \left(\sqrt{\frac{1}{5} \left(21-2 \sqrt{65}\right)}\right) \csc \left(2 \arctan \left(\sqrt{\frac{1}{5} \left(21-2 \sqrt{65}\right)}\right)\right) \\
        &\approx 0.96642i \\
        g_{1, 3} &= -\frac{\sqrt{2 \sqrt{10}+5} \pi }{2 \sqrt[4]{10}} \approx -2.97256 \\
        g_{2, 3} &= \frac{\sqrt{2 \sqrt{10}-5} \pi }{2 \sqrt[4]{10}} \approx 1.01661
    \end{align*}
    \item \(C_2\):
    \begin{align*}
        r_{3, 2} &= -i \arctan \left(\sqrt{\frac{13}{3}}\right) \approx -1.12296  i \\
        r_{3, 1} &= i \arctan \left(2 \sqrt{\frac{2}{3}}\right) \approx 1.02133 i
    \end{align*}
    \item \(Q_1\):
    \begin{align*}
        g_{1, 2} &= \frac{8 i \arctan \left(\sqrt{\frac{47}{5}-2 \sqrt{\frac{26}{5}}}\right)}{\sqrt{47-2 \sqrt{130}}}
        \approx 1.86077 i \\
        g_{2, 2} &= i \sqrt{\frac{10}{13}} \arctan \left(\sqrt{\frac{47}{5}-2 \sqrt{\frac{26}{5}}}\right) \csc \left(2 \arctan \left(\sqrt{\frac{47}{5}-2 \sqrt{\frac{26}{5}}}\right)\right) \\
        &\approx 1.33183 i \\
        g_{1,1} &= -i \arctan \left(\sqrt{\frac{11}{5}}\right)
        \approx -0.97760 i
    \end{align*}
    \item \(C_1\):
    \begin{align*}
        r_{1, 0} &= \frac{8 i \arctan \left(\frac{1}{2} \sqrt{\frac{1}{2} \left(37-3 \sqrt{15}\right)}\right) \csc \left(2 \arctan \left(\frac{1}{2} \sqrt{\frac{1}{2} \left(37-3 \sqrt{15}\right)}\right)\right)}{3 \sqrt{5}} \\
        & \approx 1.47959 i \\
        r_{2, 0} &= 2 i \sqrt{\frac{13}{37-3 \sqrt{15}}} \arctan \left(\frac{1}{2} \sqrt{\frac{1}{2} \left(37-3 \sqrt{15}\right)}\right) \\
        & \approx 1.51613 i
    \end{align*}
\end{itemize}

\section{Appendix: quantum speed limit for time dependent Hamiltonians} \label{sec_app_qsl}
It has long been known that for quantum dynamics, unitary or non-unitary, are constrained by a quantum speed limit (QSL)~\cite{Campo2013a,Taddei2013a,Toth2014a,Nishiyama2026a} where the fidelity or other distance measures can only change by a amount bounded by the dynamical map. To make the derivation self-contained, we include a short sketch of the proof for the quantum speed limit for time-dependent systems evolving either unitarily or non-unitarily in our notation for this work.

Let \(|\psi_r \rangle\) be a given reference state, the QSL was originally derived [?] specifying the minimum time \(T\), \(\int_0^T\Delta H(s)ds\) as resource, to change fidelity or some other distance measure by a given amount.

Denote \(f = \tr(\rho \Pi_r)\) as fidelity, \(\Pi_r = |\psi_r \rangle\langle \psi_r|\) is a projector, and \(\varphi = \arccos \sqrt{f}\).
\begin{align}
    \partial_t \varphi &= -\dot{f} / 2\sqrt{f(1-f)},\\
    &= -\tr(\dot{\rho} \Pi_r) / 2\sqrt{f(1-f)}
\end{align}
since \(\tr(\rho)=1\), \(\tr(\dot\rho)=0\), so \(\tr(\dot{\rho} \Pi_r) = \tr[\dot{\rho} (\Pi_r-f)] \equiv \tr(\dot{\rho} \tilde\Pi_r)\).

Denote the SLD operator for \(\dot{\rho}\) as \(L_t\), \(\dot{\rho}=\{ \rho, L_t \}/2\),
\begin{align}
    \partial_t \varphi &= -\frac{\tr\left[\{ \rho, L_t \} \tilde\Pi_r\right]}{ 4\sqrt{f(1-f)} } \\
    &= -\frac{\tr\left[\rho L_t \tilde\Pi_r + \rho \tilde\Pi_r L_t \right] / 2}{ 2\sqrt{f(1-f)} }
\end{align}
using the Cauchy-Schwarz inequality for Hermitian \(A, B\),
\begin{align}
    \left|\tr(\rho \{A, B\})/2\right| &\leq \sqrt{\tr\left(\rho A^2 \right)} \sqrt{\tr\left(\rho B^2 \right)}
\end{align}
We have,
\begin{align}
    |\partial_t \varphi|
    &= \frac{\left|\tr\left[\rho L_t \Pi_r + \rho \Pi_r L_t \right]\right| / 2}{ 2\sqrt{f(1-f)} } \\
    &\leq \frac{\sqrt{\tr\left(\rho L_t^2 \right)} \sqrt{\tr\left(\rho \tilde\Pi_r^2 \right)} }{ 2\sqrt{f(1-f)} }
\end{align}
\(\tr\left(\rho L_t^2 \right)\) is the QFI wrt the time as formal parameter \(\mathcal{F}\), and
\begin{align}
    \tr\left(\rho \tilde\Pi_r^2 \right) &= \tr\left[\rho (\Pi_r + f^2 - 2f\Pi_r) \right] \\
    &= f + f^2 - 2 f^2 = f(1-f)
\end{align}
so, for \(\varphi = \arccos \sqrt{f}\), the QSL can be written as
\begin{align}
    |\partial_t \varphi| &\leq \sqrt{\mathcal{F}}/2
\end{align}
or in integration form,
\begin{align}
    |\varphi(T) - \varphi(0)| &\leq \int_0^T ds \sqrt{\mathcal{F}(s)}/2 \equiv \Gamma
\end{align}

For unitary evolution with pure states, the expression above reduces to
\begin{align}
    f = \tr(\rho \Pi_r) = \langle \psi_r | \psi(t) \rangle\langle \psi(t) | \psi_r  \rangle = |\langle \psi(t) | \psi_r  \rangle|^2,
\end{align}
\(\varphi = \arccos \sqrt{f} = \arccos |\langle \psi(t) | \psi_r  \rangle|\), and the QFI for pure state
\begin{align}
    \sqrt{\mathcal{F}}/2 &= \sqrt{4 \left[\langle \partial_t \psi | \partial_t \psi \rangle - |\langle \partial_t \psi | \psi \rangle|^2\right]}/2 \\
    &=\sqrt{\langle \psi | H^2 | \psi \rangle - |\langle  \psi | H | \psi \rangle|^2} = \Delta H
\end{align}
so
\begin{align}
    |\varphi(T) - \varphi(0)| &\leq \int_0^T ds \Delta H(s) \equiv \Gamma
\end{align}
put in QSL time form,
\begin{align}
    & |\varphi(T) - \varphi(0)|/T \leq \frac{1}{T}\int_0^T ds \Delta H(s) \\
    & T\geq \frac{|\varphi(T) - \varphi(0)|}{\frac{1}{T}\int_0^T ds \Delta H(s)} \equiv \frac{|\varphi(T) - \varphi(0)|}{\overline{\Delta H}} \equiv T_{QSL}.
\end{align}

\bibliographystyle{iopart-num}
\bibliography{refs}

\end{document}